\documentclass[12pt]{article}

\usepackage{amssymb}

\newcommand{\Z}{\mathbb{Z}}

\newcommand{\R}{\mathbb{R}}
\newcommand{\C}{\mathbb{C}}

\newcommand{\Tr}{{\rm Tr}~}

\newcommand{\beq}{\begin{equation}}
\newcommand{\eeq}{\end{equation}}
\newcommand{\beqa}{\begin{eqnarray}}
\newcommand{\eeqa}{\end{eqnarray}}
\newcommand{\beqs}{\begin{displaymath}}
\newcommand{\eeqs}{\end{displaymath}}
\newcommand{\beqas}{\begin{eqnarray*}}
\newcommand{\eeqas}{\end{eqnarray*}}

\title{Deformation Quantisation of Gravity}
\author{\large Frank Antonsen \\University of Copenhagen \\Niels Bohr 
Institute\\ Blegdamsvej 17\\ DK-2100 Copenhagen {\O}\\ Denmark
}

\begin{document}
\maketitle

\begin{abstract}
We study the deformation (Moyal) quantisation of gravity in both the ADM and 
the Ashtekar approach. It is shown, that both can be treated, but lead to 
anomalies. The anomaly in the case of Ashtekar variables, however, is merely
a central extension of the constraint algebra, which can be ``lifted''.\\
Finally we write down the equations defining physical states and comment on 
their physical content. This is done by defining a loop representation. We 
find a solution in terms of a Chern-Simons state, whose Wigner function then
becomes related to BF-theory. This state exist even in the absence of a 
cosmological constant but only if certain extra conditions are imposed. 
Another
solution is found where the Wigner function is a Gaussian in the momenta.\\
Some comments on ``quantum gravity'' in lower dimensions are also made.
\end{abstract}

\section{Introduction}
This paper is based on \cite{wdef}, where it was shown in more detail
how to perform a deformation quantisation of constrained systems. Among
other things it was shown how classical second class constraints could be
turned into first class quantum constraints. It was also shown how, in certain
cases, simple kind of anomalous contributions could be ``lifted'', i.e., one
could find quantum constraints which did not have this anomalous contribution
to their Moyal algebra, but instead might have a singular naive classical 
limit, $\hbar\rightarrow 0$. Consequently, the correct classical limit is
obtained as the principal part of the $\hbar\rightarrow 0$ limit.\\
We will give a short summary of the techniques here, and then expand on the
resulting treatment of gravitation sketched upon in \cite{wdef}.\\
Deformation quantisation, \cite{deform}, consists essentially in replacing 
the classical
Poisson bracket, $\{\cdot,\cdot\}_{\rm PB}$ by a new bracket known as the
{\em Moyal bracket}
\beq
	[f,g]_M := i\hbar \{f,g\}_{\rm PB} + O(\hbar^2)
\eeq
where $f,g$ are functions on the classical phase-space $\Gamma$. Hence,
deformation quantisation keeps the classical phase-space but endows it with
a new, deformed bracket. For a flat phase-space, i.e., $\Gamma\simeq \R^{2n}$,
the Moyal bracket is essentially unique, \cite{Tzan,DerVer}. It is given by
\beq
	[f,g]_M = 2i f\sin\left(\frac{1}{2}\hbar\bigtriangleup\right) g
\eeq
where $\bigtriangleup$ is the bidifferential operator giving the Poisson
bracket, i.e.,
\beq
	f\bigtriangleup g := \{f,g\}_{\rm PB}
\eeq
Consequently,
\beqa
	[f,g]_M &=& i\sum_{n=0}^\infty \frac{(-1)^n\hbar^{2n+1}}{4^n (2n+1)!}
	\sum_{k=0}^n (-1)^k\left(\begin{array}{c}n\\k\end{array}\right)
	\frac{\partial^n f}{\partial q^{n-k}\partial p^k}\frac{\partial^n g}
	{\partial p^{n-k}\partial q^k}\\
		&:=& i\sum_{n=0}^\infty \hbar^{2n+1}\omega_{2n+1}(f,g)
\eeqa
We will also introduce the {\em anti-Moyal bracket}
\beq
	[f,g]_M^+ := 2 f\cos\left(\frac{1}{2}\hbar\bigtriangleup\right)g
\eeq
Both of these can be written in terms of a {\em twisted product} $*$ as
\beq
	[f,g]_M = f*g-g*f \qquad [f,g]_M^+ = f*g+g*f
\eeq
This twisted product is a deformation of the usual product of functions,
\beq
	f*g = fg+O(\hbar)
\eeq
and comes from the {\em Weyl map} associating functions $A_W(q,p)$ on the
classical phase-space to operators $\hat{A}$ acting on $L^2(Q)$ where $Q$ 
is the coordinate manifold. This map is given by, \cite{Dahl,wlie}
\beq
	A_W(q,p) := \int e^{iuq-ivp} {\rm Tr}(\Pi(u,v)\hat{A}) dudv
\eeq
where $\Pi(u,v)$ is a translation operator on $T^*Q$. When $Q=\R^n$, it
is given explicitly by
\beq
	\Pi(u,v)= e^{iu\hat{q}-iv\hat{p}}
\eeq
which form a ray representation of the Euclidean group.
The inverse of this map is
\beq
	A^W := \int A_W(q,p)\Pi(u,v) e^{iuq-ivp} dudvdqdp
\eeq
and the twisted product is induced from the noncommutative product of operators
\beq
	(\hat{A}\hat{B})_W = A_W*B_W
\eeq
implying
\beq
	([\hat{A},\hat{B}])_W = [A_W,B_W]_M \qquad (\{\hat{A},\hat{B}\})_W
	= [A_W,B_W]_M^+
\eeq
where $\{\cdot,\cdot\}$ is the anticommutator. This relationship illuminates
the standard Heisenberg-Dirac rule 
\beq
	\{f,g\}_{\rm PB} \rightarrow \frac{1}{i\hbar}[\hat{f},\hat{g}]
\eeq
Furthermore, the Heisenberg-Dirac rule is known not always to work (the
Groenwold-van Hove no-go theorem), 
\cite{geomq,Sd}, whereas deformation quantisation always work, 
\cite{deform,Kontsevich}.\\
For constrained systems, deformation quantisation was studied, probably for
the first time, in \cite{wdef}. Here I'll just give a brief outline.\\
Consider a flat phase-space $\Gamma$ and a set of constraints $\phi_a(q,p)$.
Assume first of all that these are all first class, i.e.,
\beq
	\{\phi_a,\phi_b\}_{\rm PB} = c_{ab}^{~~c}\phi_c
\eeq
and in involution with the Hamiltonian $h$, i.e.,
\beq
	\{h,\phi_a\}_{\rm PB} = V_a^b\phi_b
\eeq
Then deformation quantisation consists in finding quantum constraints $\Phi_a$
and Hamiltonian $H$, such that
\beqa
	\left[\Phi_a,\Phi_b\right]_M &=& i\hbar c_{ab}^{~~c}\Phi_c\\
	\left[H,\Phi_a\right]_M &=& i\hbar V_a^b\Phi_b
\eeqa
By assuming $\Phi_a,H$ to be analytic functions of the deformation parameter
$\hbar$, i.e., that they can be Taylor expanded, the above Moyal brackets
define a recursive scheme for finding $\Phi_a,H$. Furthermore, when $\phi_a,h$
are at most cubic in $p,q$ or has the form of a cubic polynomial plus a 
function of only one of the canonical variables, $q$, say, then we can pick
$H=h,\Phi_a=\phi_a$.\\
If the constraints are second class, $\{\phi_a,\phi_b\}_{\rm PB} = \chi_{ab}
\neq c_{ab}^{~~c}\phi_c$, or not in involution with the Hamiltonian, $\{h,
\phi_a\}_{\rm PB}\neq V_a^b\phi_b$, then new quantum constraints and/or 
Hamiltonian, $\Phi_a, H$, can be found which are first class and in involution.
The price one has to pay is the inclusion of negative powers of $\hbar$ in the
formal power-series defining $H,\Phi_a$, it might also be necessary to allow
the structure coefficients to receive quantum corrections. I.e., one can obtain
\beqa
	\left[\Phi_a,\Phi_b\right]_M &=& i\hbar \tilde{c}_{ab}^{~~c}\Phi_c\\
	\left[H,\Phi_a\right]_M &=& i\hbar \tilde{V}_a^b\Phi_b
\eeqa
with
\beqa
	\Phi_a &=& \hbar^{-1}\Phi_a^{(-1)}+\phi_a+\hbar\Phi_a^{(1)}+...\\
	H &=& h+\hbar H^{(1)}+...\\
	\tilde{c}_{ab}^{~~c} &=& c_{ab}^{~~c} +O(\hbar)\\
	\tilde{V}_a^b &=& V_a^b+O(\hbar)
\eeqa
A similar trick can also take care of ``anomalies'' where $[\phi_a,\phi_b]_M
=i\hbar\{\phi_a,\phi_b\}_{\rm PB} +O(\hbar^3)$, at least when the quantum
correction is a constant (i.e., the constraint algebra gets centrally extended
when one naively uses the classical constraints in the Moyal brackets).
See \cite{wdef} for more details. It should be noticed that Hamachi
independently has arrived at a similar ``quantum smoothening'' of anomalies,
\cite{Hamachi}, albeit in somewhat simpler situations only applicable to
toy models, at least at the present state, but with a much higher level of
mathematical rigour.\\
The classical condition picking out physical states is simply $\phi_a(q,p)=0,
\forall a$. In the standard Dirac quantisation picture, this would get replaced
by $\hat{\phi}_a|\psi\rangle=0, \forall a$, where $\hat{\phi}_a$ is some
operator corresponding to $\phi_a$, i.e., satisfying the same algebra but with
Poisson brackets replaced by $(i\hbar)^{-1}$ times commutators.\\  
In an improved version of the
Dirac condition, the BRST-condition, one imposes instead $\hat{\Omega}|\psi
\rangle=0$ for a state $|\psi\rangle$ and $[\hat{\Omega},\hat{A}]=0$ for an
observable $\hat{A}$. Here $\hat{\Omega}$ is the BRST-operator, $\hat{\Omega}
= \eta^a\hat{\phi}_a+...$ where $\eta^a$ are ghosts and where the commutator is
understood to be graded appropriately, \cite{brst}. \footnote{It ought 
to be emphasised
that BRST is also subject to the Groenwold-van Hove no-go theorem, but it
ameliorates the problem by using the underlying cohomological structure better
that usual Dirac quantisation. The BRST symmetry is after all a classical
symmetry, and the transition $\Omega\rightarrow\hat{\Omega}$ is subject to more
or less the same problems as the naive transition $(q,p)\rightarrow (\hat{q},
\hat{p})$. In both cases, the step of quantisation can probably be defined 
rigorously only through a deformation quantisation procedure in general.}\\
Deformation quantisation comes with another alternative. In \cite{wdef} it
was proposed to use
\beq
	[\Phi_a,W]_M^+=0 ~~,~~\forall a \label{eq:phys}
\eeq
to pick out physical states $W$ (Wigner functions), and similarly for other
observables $A$, $[\Phi_a,A]_M=0, \forall a$. The Wigner-Weyl-Moyal formalism
treats observables, states and transitions on an equal footing. The
semi-classical limit of (\ref{eq:phys}) is
\beq
	\phi_aW^{(0)} = 0
\eeq
where $\phi_a, W^{(0)}$ are the $\hbar\rightarrow 0$ limits of $\Phi_a,W$
respectively. Thus
\beq
	W^{(0)} \propto \prod_a \delta (\phi_a)
\eeq
i.e., $W^{(0}$ vanishes away from the constraint surface, at least in a 
distributional sense. We can also write this as
\beq
	{\rm supp}~ W^{(0)} \subseteq \bigcap_a\ker\phi_a
\eeq
The replacement of Poisson brackets by Moyal ones has been used by Strachan,
Takasaki, Pleba\'{n}sky and coworkers to study self-dual gravity and 
Yang-Mills theory, see e.g. \cite{selfdual}.\\
A few comments are in order. First, the replacement of Poisson brackets by
Moyal ones implies that the corresponding ``gauge'' transformations acquire
quantum modifications. If the classical constraints are denoted by
$\phi_a$ they
generate (infinitesimal) ``gauge'' transformations $\delta_\omega f:= 
\{\omega^a\phi_a,f\}_{\rm PB}$, the corresponding quantum version is
\beq
	\delta_\omega F := [\omega^a\Phi_a,F]_M = i\hbar\{\omega^a\phi_a,F
	\}_{\rm PB} + \mbox{other terms}
\eeq
which a priori differs from the classical expression. The discrepancy between
the classical and the quantum ``gauge'' transformations show up in higher
order derivatives, which seems to suggest that the quantum transformations
are ``larger'', i.e., slightly less local than their classical counterparts.\\
Another point to check is whether the space of physical quantities is 
invariant under such transformations. Consider thus an element $A$ satisfying
$[\Phi_a,A]_M^+=0, ~\forall a$, when then wants to prove that a ``gauge''
transformation does not take us away from this subspace, i.e., $\delta_\omega
[\Phi_a,A]_M^+=0, ~\forall a$. We get
\beqa
	\delta_\omega ([\Phi_a,A]_M^+) &=& [\delta_\omega\Phi_a,A]_M^++
	[\Phi_a,\delta_\omega A]_M^+\nonumber\\
	&=& [\omega^b c_{ba}^{~~c}\Phi_c,A]_M^+ + [\Phi_a,[\omega^b\Phi_b,A]_M
	]_M^+
\eeqa
the first term is zero provided $c_{ab}^{~~c}$ has vanishing Moyal brackets
with $A$, e.g., if the structure coefficient is independent of the phase-space
variables. The second term is also zero, as one can see by noting that
$\Phi_a*A=-A*\Phi_a, ~\forall a$, since we can then rewrite the second term
as
\beq
	\delta_\omega[(\Phi_a,A]_M^+) = \omega^b[[\Phi_a,\Phi_b]_M,A]_M^+ =0
\eeq
Even if $c_{ab}^{~~c}$ depends upon the phase-space variables, as is the case,
for instance, in gravity, then $\delta_\omega[\Phi_a,A]_M^+$ still vanishes
since we have in general
\beq
	\delta_\omega([\Phi_a,A]_M^+) = \omega^b\left([[\Phi_b,\Phi_a]_M,A]_M^+
	+[[\Phi_a,\Phi_b]_M,A]_M^+\right) = 0
\eeq
Hence the condition $0=[\Phi_a,A]_M^+$ is a consistent quantum analogue of
$\phi_a=0$, as we had anticipated.\\
It is argued in \cite{def} that the approach put forward here is in fact
compatible with the classical BRST-symmetry, and moreover, that the
entire deformation quantisation procedure can be expressed in geometrical
terms (through a sheaf over the real axis).\\
We will now apply this formalism to gravity.

\section{ADM Variables}
In the approach due to Arnowit, Desser and Misner, 
\cite{ADM}, one splits up the metric as
\beq
	g_{\mu\nu} = \left(\begin{array}{cc} N & N_i\\
	N_j & g_{ij}\end{array}\right)
\eeq
where $N$ is known as the lapse function and $N_i$ as the
shift vector -- these are Lagrange multipliers just like $A_0^a$ for
the Yang-Mills case. The proper canonical variables
are then the 3-metric $g_{ij}$ (again, for Yang-Mills theory it is the 3-vector
$A_i^a$) and its conjugate momentum $\pi^{ij}$. Hence the spacetime manifold
has to be globally hyperbolic, $M\simeq\Sigma\times\R$, where 
$\Sigma$ is a spatial hypersurface. Consequently, the ADM-approach
tells us that spacetime is to be considered not as a single four-dimensional
entity but rather as a foliation by spatial hypersurfaces, i.e.,
$M$ is the family $\{\Sigma_t\}_{t\in\R}$ where $\Sigma_t\simeq \Sigma, \forall
t\in \R$. The choice of ``time'' $t$ is a gauge fixing, as are the choices of
$N,N_i$.
\\
The action can be written as
\beq
	S = \int (\dot{g}_{ij}\pi^{ij} - N{\cal H}_\perp-N^i{\cal H}_i) dx
\eeq
where ${\cal H}_\perp,{\cal H}_i$ are constraints depending only on $g_{ij}$
and $\pi^{ij}$ -- the equations of motion
of $N,N_i$ gives the constraints ${\cal H}_\perp={\cal H}_i=0$.
\beqa
	{\cal H}_\perp(x) &=& g^{-1/2}(\frac{1}{2}\pi^2-\pi^i_j\pi^j_i)+
	\sqrt{g}R := G_{ijkl}\pi^{ij}\pi^{kl}+\sqrt{g}R\\
	{\cal H}_i(x) &=& -2D_j\pi^j_i
\eeqa
with $g_{ij}$ the 3-metric, $\pi^{ij}$ its conjugate momentum,
\beq
	\{g_{ij}(x),\pi^{kl}(x')\}_{\rm PB} = \frac{1}{2}(\delta^k_i\delta^l_j
	+\delta^k_j\delta^l_i)\delta(x,x'),
\eeq
$R$ the curvature scalar of $g_{ij}$ (i.e., the three dimensional one)
and $g$ the determinant of the 3-metric. The first constraint is known as
the Hamiltonian one, and the last, the 
${\cal H}_i$, as the diffeomorphism one. The algebra is
\beqa
	\{{\cal H}_\perp(x),{\cal H}_\perp(x')\}_{\rm PB} &=& (g^{ij}(x)
	{\cal H}_j(x)+g^{ij}(x'){\cal H}_j(x'))\delta_{,i}(x,x')\\
	\{{\cal H}_\perp(x),{\cal H}_i(x')\}_{\rm PB} &=& {\cal H}_\perp(x)
	\delta_{,i}(x,x')\\
	\{{\cal H}_i(x),{\cal H}_j(x')\}_{\rm PB} &=& {\cal H}_i(x')\delta_{,j}
	(x,x')+{\cal H}_j(x)\delta_{,i}(x,x')
\eeqa
where the subscript $\delta_{,i}$ denotes the partial derivative with respect
to $x^i$. The convention is the standard one in which $\delta(x,x')$ is a
scalar in the first argument and a density in the second (the curved
spacetime Dirac $\delta$ has a $g^{-1/2}$ in it).
\\
The algebra of ${\cal H}_i$ is ${\rm diff}(\Sigma)$, the algebra of
spatial diffeomorphism, i.e., the symmetry given by this first class constraint
is the diffeomorphism symmetry of $\Sigma$. The Hamiltonian constraint
generates ``motion'' away from one spatial slice $\Sigma_t$ to another
$\Sigma_{t'}, t'\geq t$, i.e., the time evolution (with the given definition of
time coordinate) of the three-manifold.
\\
It is important to notice that the structure coefficients of the constraint
algebra depend upon the phase-space variables (the 3-metric). Consequently, a
naive canonical quantisation is very troublesome; when $g_{ij},\pi^{ij}$
becomes operators the structure coefficients will no longer commute with
the constraints, so in which order is one to write down the quantum
constraint algebra, are the $g_{ij}$ to stand to the left or the right of
the constraints? In order to ensure that time evolution does not take
one away from the constraint surface, one has to demand that the constraints
stand to the right of the structure coefficients on the right hand side of
the constraint algebra, and it is this which makes a standard 
canonical quantisation of gravity so difficult.
\\
Deformation quantisation does not care about such problems.\\
The algebra of the diffeomorphism constraint will not be deformed as their 
form is ${\cal H}_i \sim \partial\pi + g^2\pi$ and thus has $\omega_3\equiv 
0$. The Hamiltonian constraint, however, has as well a $g\pi^2$ as a $g^2\pi$
term, and will consequently not have vanishing $\omega_3$. We should thus
expect the algebraic relations involving ${\cal H}_\perp$ to receive $\hbar^3$
corrections (but no higher order corrections since no higher powers of $\pi$
are present). This is precisely what we find. Moreover, the Christoffel
symbols and the $\sqrt{g}$ contain, in a Taylor series, the metric to infinite
order, whence we should expect infinite order equations to turn up.\\
In fact, gravity in the ADM approach with constraints ${\cal H}_\perp,{\cal 
H}_i$ is anomalous upon a deformation quantisation in the sense that
\beqas
	\left[{\cal H}_\perp(x),{\cal H}_\perp(x')\right]_M 
	&=& i\hbar\{{\cal H}_\perp(x),{\cal H}_\perp(x')\}_{\rm PB}+
	i\hbar^3 k(x,x')\\
	\left[{\cal H}_\perp(x),{\cal H}_i(x')\right]_M 
	&=& i\hbar\{{\cal H}_\perp(x),{\cal H}_i(x')\}_{\rm PB} +
	i\hbar^3 k_i(x,x')
\eeqas
whereas the spatial diffeomorphism subalgebra generated by the ${\cal H}_i$
is non-anomalous.\\
A straightforward computation yields
\beq
	k(x,x') = -\frac{1}{8}\left((\Xi_{mnab}^{ijkl}\pi^{ab})(x)
	{\cal G}^{mn}_{ijkl}(x') - (x\leftrightarrow x')\right)
\eeq
with
\beqa
	\Xi_{mnab}^{ijkl} &\equiv & \frac{\delta^3{\cal H}_\perp}{\delta
	g_{ij}\delta g_{kl}\delta\pi^{mn}}\\
	&=&G_{mnab}(g^{ik}g^{jl}-\frac{1}{2}g^{ij}g^{kl})
	-\frac{1}{2}g^{ij}\left(g_{nb}\delta^k_m\delta^l_a+g_{ma}\delta^k_n
	\delta^l_b-\right.\nonumber\\
	&&\qquad\left.g_{ab}\delta^k_m\delta^l_n-g_{mn}\delta^k_a\delta^l_b
	+g_{bn}\delta^l_m\delta^k_a+g_{am}\delta^l_n\delta^k_b\right)\\
	{\cal G}^{mn}_{ijkl} &\equiv& \frac{\delta^3{\cal H}_\perp}{\delta
	\pi^{ij}\delta\pi^{kl}\delta g_{mn}}\\ 
	&=&-\frac{1}{2}g^{mn}G_{ijkl}+g^{-1/2}\left(g_{jl}(
	\delta^m_i\delta^n_k+\delta^m_k\delta^n_i)+g_{ik}(\delta^m_j\delta^n_l
	+\delta^m_l\delta^n_j)-\right.\nonumber\\
	&&\qquad\left. g_{kl}\delta^m_i\delta^n_j-g_{ij}\delta^m_k\delta^n_l
	\right)
\eeqa
One should note that $[{\cal H}_\perp,k]_M\neq 0$ hence we get an anomaly which
is {\em not} a central extension of the original algebra. Explicitly
\beqa
	[{\cal H}_\perp,k]_M = i\hbar\{{\cal H}_\perp,k\}_{\rm PB} +i\hbar^3
	\frac{3}{4}\Xi^{ijkl}_{mnab}\pi^{ab}
	\frac{\delta^3 k}{\delta\pi^{ij}\delta g_{kl}\delta
	g_{mn}}\neq 0
\eeqa
For the ADM constraints, the structure coefficients depend on the
fields, consequently the anomaly too depends upon $(g,\pi)$.
\\
Similarly, the relation mixing ${\cal H}_\perp$ and ${\cal H}_i$ receives a
$\hbar^3$ correction of the form
\beqs
	k_i(x,x')\equiv
	-\frac{1}{8}\frac{\delta^3{\cal H}_\perp(x)}{\delta\pi^{jk}\delta
	\pi^{lm}\delta g_{ab}}\frac{\delta^3{\cal H}_i(x')}{\delta g_{jk}\delta
	g_{lm}\delta\pi^{ab}}
\eeqs
which one easily finds to be 
\beq
	k_i(x,x') = -\frac{1}{4}{\cal G}_{jklm}^{pq}\Upsilon_{ipq}^{jklm}
\eeq
with
\beqa
	\frac{\delta^3{\cal H}_i(x)}{\delta g_{jk}(x')\delta g_{lm}(x'')
	\delta\pi^{ab}(y)} &=& 2\Upsilon_{iab}^{jklm}(x,x',x'')\delta(x,y)
	\nonumber\\
	&=& \delta^c_{(a}\delta^n_{b)}\delta(x,y)\left\{\delta^l_{(n}
	\delta^m_{i)}\delta(x,x'')\left(-g^{rk}\Gamma^j_{rc}\delta(x,x')
	+\right.\right.\nonumber\\
	&&\left.\frac{1}{2}g^{rs}\left(\delta^j_{(r}\delta^k_{s)}\partial_c
	+\delta^j_{(s}\delta^k_{c)}\partial_r-\delta^j_{(c}\delta^k_{r)}
	\partial_s\right)\delta(x,x')\right)+\nonumber\\
	&&\delta^j_{(n}\delta^k_{i)}\delta(x,x')\left(-g^{rm}\Gamma^l_{rc}
	\delta(x,x'')
	+\right.\nonumber\\
	&&\left.\frac{1}{2}g^{rs}\left(\delta^l_{(r}\delta^m_{s)}\partial_c
	+\delta^l_{(s}\delta^m_{c)}\partial_r-\delta^l_{(c}\delta^m_{r)}
	\partial_s\right)\delta(x,x'')\right)-\nonumber\\
	&&g_{ni}\delta(x,x'')\left[g^{rl}g^{km}\Gamma^j_{rc}-g^{rk}
	g^{jm}\Gamma^l_{rc}+\right.\nonumber\\
	&&\frac{1}{2}g^{rk}g^{js}\left(\delta^l_{(r}
	\delta^m_{s)}\partial_c+\delta^l_{(s}\delta^m_{c)}\partial_r
	-\delta^l_{(r}\delta^m_{c)}\partial_s\right)-\nonumber\\
	&&\left.\left.\frac{1}{2}g^{rl}g^{sm}\left(\delta^j_{(r}
	\delta^k_{s)}\partial_c+\delta^j_{(s}\delta^k_{c)}\partial_r
	-\delta^j_{(r}\delta^k_{c)}\partial_s\right)\right]\delta(x,x')\right\}
	+\nonumber\\
	&&(c\rightarrow r, n\rightarrow c, r\rightarrow n)
\eeqa
The spatial diffeomorphism subalgebra spanned by ${\cal H}_i$ does not receive
any quantum corrections since the constraints are only linear in the 
momentum.\\
We have relied on the following
\beqas
	\frac{\delta g(x)}{\delta g_{ij}(x')} &=& g g^{ij}\delta(x,x')\\
	\frac{\delta g^{ij}(x)}{\delta_{kl}(x')} &=& -g^{ik}g^{jl}
	\delta(x,x')\\
	\frac{\delta\Gamma^i_{jk}(x)}{\delta g_{mn}(x')} &=& -g^{im}
	\Gamma^n_{jk}\delta(x,x')+\frac{1}{2}g^{il}\left(\delta^m_{(j}
	\delta^n_{l)}\partial_l+\delta^m_{(k}\delta^n_{l)}\partial_j
	-\delta^m_{(j}\delta^n_{k)}\partial_l\right)\delta(x,x')
\eeqas
the first of which can be found in \cite{Nakahara}, and the last is a 
straightforward consequence of the first two relations.

\subsection{Physical States}
The set of physical states are defined as the functions $W$ satisfying the 
two infinite order functional differential equations
\beqa
	0 &=& [{\cal H}_\perp,W]_M^+\\
	0 &=& [{\cal H}_i,W]_M^+
\eeqa
these are infinite order since the Christoffel symbols (and hence the covariant
derivative and the curvature scalar) has an inverse metric in them, similarly
the supermetric $G_{ijkl}$ too has an inverse metric inside. Thus the 
constraints are not polynomial in the metric, but instead ``meromorphic''.
\\
Written out more explicitly, the physicality conditions read 
\beqa
	0 &=& 2{\cal H}_\perp W +\sum_{k=1}^\infty (-1)^k 2^{-2k-1}\frac{
	\hbar^{2k}}{(2k)!}
	\left(\frac{\delta^{2k}{\cal H}_\perp}{\delta g_{i_1j_1}...\delta 
	g_{i_{2k-1}j_{2k-1}}\delta g_{mn}}\frac{\delta^{2k}W}{\delta
	\pi^{i_1j_1}...\delta\pi^{i_{2k-1}j_{2k-1}}\delta\pi^{mn}}-
	\right.\nonumber\\
	&&\qquad 2k
	\frac{\delta^{2k}{\cal H}_\perp}{\delta g_{i_1j_1}...\delta 
	g_{i_{2k-1}j_{2k-1}}\delta\pi^{mn}}
	\frac{\delta^{2k}W}{\delta\pi^{i_1j_1}...\delta\pi^{i_{2k-1}j_{2k-1}}
	\delta g_{mn}}+\nonumber\\
	&&\qquad\left.
	k(2k-1)\frac{\delta^{2k}{\cal H}_\perp}{\delta g_{i_1j_1}
	...\delta g_{i_{2k-2}j_{2k-2}}\delta\pi^{i_{2k-1}j_{2k-1}}
	\delta\pi^{mn}}\frac{\delta^{2k}W}{\delta\pi^{i_1j_1}...
	\delta\pi^{i_{2k-2}j_{2k-2}}\delta g_{i_{2k-1}j_{2k-1}}\delta
	g_{mn}}\right)\nonumber\\
	&&\\
	0 &=& 2{\cal H}_i W + \sum_{k=1}^\infty (-1)^k 2^{-2k-1}
	\frac{\hbar^{2k}}{(2k)!}
	\left(\frac{\delta^{2k}{\cal H}_i}{\delta g_{i_1j_1}...\delta 
	g_{i_{2k}j_{2k}}}\frac{\delta^{2k} W}{\delta\pi^{i_1
	j_1}...\delta\pi^{i_{2k}j_{2k}}}-\right.\nonumber\\
	&&\qquad\left.
	2k\frac{\delta^{2k}{\cal H}_i}{\delta g_{i_1j_1}...\delta 
	g_{i_{2k-1}j_{2k-1}}\delta\pi^{i_{2k}j_{2k}}}
	\frac{\delta^{2k}W}{\delta\pi^{i_1j_1}...\delta\pi^{i_{2k-1}j_{2k-1}}
	\delta g_{i_{2k}j_{2k}}}\right)
\eeqa
Since the constraints for gravity in the ADM formalism are 
non-polynomial the equations defining the physical state space become infinite
order. If one assumes the Wigner function to be analytic in $\hbar$, one
can Taylor expand it $W=\sum_{n=0}^\infty \hbar^n W_n$, and arrive at the
following recursive formulas for the $n$'th order coefficients, $W_n$
\beqa
	0 &=& {\cal H}_\perp W_0 = {\cal H}_i W_0\\
	0 &=& 2{\cal H}_\perp W_N+\sum_{k=1}^{[N/2]} (-1)^k2^{-2k-1}
	\frac{1}{(2k)!}
	\left(\frac{\delta^{2k}{\cal H}_\perp}{\delta g_{i_1j_1}...
	\delta g_{i_{2k-1}j_{2k-1}}\delta\pi^{mn}}
	\frac{\delta^{2k}}{\delta\pi^{i_1j_1}...\delta\pi^{i_{2k-1}j_{2k-1}}
	\delta g_{mn}}-\right.\nonumber\\
	&&\qquad\left. 2k\frac{\delta^{2k}{\cal H}_\perp}{\delta g_{i_1j_1}
	...\delta g_{i_{2k-2}j_{2k-2}}\delta\pi^{i_{2k-1}j_{2k-1}}
	\delta\pi^{mn}}\frac{\delta^{2k}}{\delta\pi^{i_1j_1}...
	\delta\pi^{i_{2k-2}j_{2k-2}}\delta g_{i_{2k-1}j_{2k-1}}\delta
	g_{mn}}\right) W_{N-2k}\nonumber\\
	&&\\
	0 &=& 2{\cal H}_iW_N+\sum_{k=1}^{[N/2]}(-1)^k 2^{-2k-1}\frac{1}{(2k)!}
	\frac{\delta^{2k}{\cal H}_i}{\delta g_{i_1j_1}...\delta 
	g_{i_{2k-1}j_{2k-1}}\delta\pi^{mn}}
	\frac{\delta^{2k}W_{N-2k}}{\delta\pi^{i_1j_1}...
	\delta\pi^{i_{2k-1}j_{2k-1}}\delta g_{mn}}
\eeqa
where $k,N\geq 1$. 
\\
From this we get that the $W_{2n}$ decouple from the $W_{2n+1}$. Explicitly,
\beqa
	W_0 &\propto & \delta({\cal H}_1)\delta({\cal H}_2)\delta({\cal H}_3)
	\delta({\cal H}_\perp) := \delta^4({\cal H})\\
	W_1 &\propto & \delta^4({\cal H})\\
	W_2 &=& \frac{1}{16{\cal H}_\perp}\left(\frac{\delta^2{\cal H}_\perp}
	{\delta g\delta\pi}\frac{\delta^2W_0}{\delta \pi\delta g}
	-2\frac{\delta^2{\cal H}_\perp}{\delta\pi^2}\frac{\delta^2W_0}{\delta
	g^2}\right)\\
	&=& \frac{1}{16{\cal H}_i}\frac{\delta^2{\cal H}_i}{\delta g\delta\pi}
	\frac{\delta^2W_0}{\delta \pi\delta g}
\eeqa
etc., where we have suppressed the indices on $g,\pi$. 
From the two expressions for $W_2$ we get another equation for $W_0$,
namely
\beq
	{\cal H}_i\left(\frac{\delta^2{\cal H}_\perp}
	{\delta g\delta\pi}\frac{\delta^2W_0}{\delta \pi\delta g}
	-2\frac{\delta^2{\cal H}_\perp}{\delta\pi^2}\frac{\delta^2W_0}{\delta
	g^2}\right) = {\cal H}_\perp 
	\frac{\delta^2{\cal H}_i}{\delta g\delta\pi}
	\frac{\delta^2W_0}{\delta \pi\delta g}
\eeq
This can be written as a condition on ${\cal W}_0=\frac{\delta W_0}{\delta g}$.
We get
\beq
	{\cal F}_i(g,\pi) \frac{\delta{\cal W}_0}{\delta \pi} - 
	{\cal G}_i(g,\pi)\frac{\delta{\cal W}_0}{\delta g} =0 \label{eq:int}
\eeq
where
\beqas
	{\cal F}_i &=& {\cal H}_i\frac{\delta^2{\cal H}_\perp}{\delta g\delta
	\pi}-{\cal H}_\perp\frac{\delta^2{\cal H}_i}{\delta\pi\delta g}\\
	{\cal G}_i &=& 2{\cal H}_i\frac{\delta^2{\cal H}_\perp}{\delta\pi^2}
\eeqas
Consequently, the equation (\ref{eq:int}) is integrable provided
\beq
	\frac{\delta{\cal F}_i}{\delta g}=-\frac{\delta{\cal G}_i}{\delta\pi}
\eeq
But this cannot be the case, since the right hand side is independent of $\pi$
(${\cal H}_i$ is linear and ${\cal H}_\perp$ quadratic in $\pi$, since the
latter is already differentiated twice with respect to $\pi$ the result 
follows) whereas the left hand side isn't. Hence ${\cal W}_0$ is not an exact
form in $(g,\pi)$-space. The solutions of (\ref{eq:int}) and hence of the
equation for $W_2$ are then parametrised by the non-trivial elements of the
first cohomology class of $(g,\pi)$-space. Since this is an infinite 
dimensional space (being the cotangent bundle of Wheeler's superspace of all
3-metrics) the computation of its cohomology is highly non-trivial, and we
will not attempt it here.\\ 
The conclusion so far is then that in the ADM formalism gravity is
anomalous when one attempts a deformation quantisation, and furthermore, that
the physicality conditions are related to the cohomology classes of the 
cotangent bundle of superspace -- an infinite dimensional manifold. I have so
far not been able to lift the anomaly in this approach. Thus deformation
quantisation of gravity in the ADM-variables is a highly non-trivial
procedure, we will therefore turn to another description which shows more
promise.

\section{Ashtekar Variables}
We saw that the anomalous nature of the quantum deformed algebra of the
constraints in the ADM formalism were due to the constraints being
non-polynomial. It is therefore interesting to consider another formulation,
the Ashtekar variables \cite{Ashtekar}, where the 
constraints {\em are} polynomials. It has also been suggested, \cite{wletter},
that Wigner functions are easier to define in the Ashtekar approach than in the
ADM approach, thus hinting that the former are better suited for deformation
quantisation purposes.
\\
In four dimensions (and four dimensions only) with Lorentz signature (and 
{\em not} with Euclidean metric) we have an isomorphism between the Lorentz
algebra (which is the local gauge algebra of gravitation) and $su_2\otimes \C$,
\beq
	so_{3,1}\simeq su_2\otimes\C
\eeq
Consequently, we can consider gravitation as a complexified $su_2$ gauge
theory.
\\
In this formulation the canonical coordinates are a complex
$su_2$-connection $A_i^a$ and its momentum (a densitised dreibein/triad) 
$E_a^i$ (i.e., $E_a^i E_b^j\delta^{ab} = gg^{ij}$), 
and the constraints are
\beqa
	{\cal H} &=& F_{ij}^a E^i_bE^j_c\varepsilon^{bc}_{~~a}\\
	{\cal G}_a &=& D_iE^i_a\\
	{\cal D}_i &=& F_{ij}^aE^j_a
\eeqa
where $F_{ij}^a = \partial_i A_j^a-\partial_jA_i^a+\epsilon^a_{~bc}A_i^bA_j^c$
is the field strength (or curvature) 2-form, and $D_i$ is the gauge covariant
derivative, $F_{ij}^a\propto [D_i,D_j]$. As always, these quantities are
three dimensional objects, i.e., $i,j=1,2,3$ -- once more a globally hyperbolic
spacetime is assumed, $M\simeq \R\times
\Sigma$, where $\Sigma$ is a Cauchy surface.
\\
The first constraint, ${\cal H}$, is referred to as the 
Hamiltonian, the ${\cal G}_a$ as the Gauss and
the ${\cal D}_i$ as the diffeomorphism constraint. 
It will turn out that in these variables the anomaly is much simpler, namely
merely a central extension. In fact, for gravity in the Asthekar variables, 
the only anomalous bracket is
\beqs
	[{\cal H}(x),{\cal D}_i(x')]_M = i\hbar\{{\cal H}(x),{\cal D}_i(x')
	\}_{\rm PB} -9i\hbar^3\delta_{,i}(x,x')
\eeqs
Consequently the anomaly is a central extension and can be lifted.\\
This is seen as follows. Only the following two brackets can possibly receive
any quantum corrections, and then only to lowest order
\beqa
	\left[{\cal H}(x),{\cal H}(x')\right]_M 
	&=& i\hbar\{{\cal H}(x),{\cal H}(x')
	\}_{\rm PB}+\frac{3}{4}i\hbar^3\left(\frac{\delta {\cal H}(x)}
	{\delta A^2\delta E}\frac{\delta{\cal H}(x')}{\delta E^2\delta A}
	-(x\leftrightarrow x')\right)\nonumber\\
	&&\\
	\left[{\cal H}(x),{\cal D}_i(x')\right]_M 
	&=& i\hbar\{{\cal H}(x),{\cal D}_i(x')
	\}_{\rm PB} - \frac{3}{4}i\hbar^3\frac{\delta^3{\cal H}(x)}{\delta E^2
	\delta A}\frac{\delta^3{\cal D}_i(x')}{\delta A^2 \delta E}	
\eeqa
where we have suppressed the indices on the $A,E$. An explicit and 
straightforward computation gives
\beqa
	\omega_3({\cal H}(x),{\cal H}(x')) &=& -12 i\delta(x,x')\left(E^j_a(x)
	A_j^a(x')-E^j_a(x')A_j^a(x)\right)\\
	\omega_3({\cal H}(x),{\cal D}_m(x')) &=& 9i\delta_{,m}(x,x')
\eeqa
We notice that the first of these vanish in the sense of distributions, hence
the only quantum correction is the constant (w.r.t. the phase-space variables)
$9i\delta_{,m}(x,x')$. Consequently, the anomalous nature of gravity shows
itself in the Ashtekar variables simply in a central extension of the
constraint algebra (similar to a Schwinger term in current algebra).
\beq
	[{\cal H}(x),{\cal D}_i(x') ]_M = i\hbar\{{\cal H}(x),{\cal D}_i(x')
	\}_{\rm PB} -9i\hbar^3\delta_{,i}(x,x')
\eeq
As mentioned earlier, such central extensions can be
``lifted'' by means of a redefinition of the quantum constraints.\\
Now, one {\em can} define Ashtekar variables also in $d=2+1$ dimensions. 
Gravity in $d\leq 3$ is classically trivial, and what one gets in $d=2+1$ in
the Ashtekar approach is an $SO(2,1)$ Yang-Mills theory with flat gauge 
curvature, i.e., the constraints are \cite{Ashtekar2}
\beqs
	\epsilon^{ij}D_iE^a_j = 0 \qquad
	F_{ij}^a=0
\eeqs
Consequently, $2+1$ dimensional Ashtekar gravity is not anomalous upon a 
deformation quantisation since no higher powers of the momenta occurs and hence
$\omega_3\equiv 0$. This shows that the anomaly is very much dependent on the
dimensionality of spacetime.

\subsection{Lifting the Anomaly}
We will now attempt to find new quantum constraints $H,D_i$ such that the
anomaly vanishes (or rather, is absorbed into the redefinition of either
the constraints or the structure coefficients).\\
We will thus write
\beq
	H=\sum_{n\in\Z} \hbar^nH_n\qquad D_i=\sum_{n\in\Z}\hbar^n D_i^{(n)}
\eeq
with $H_0={\cal H}, D_i^{(0)}={\cal D}_i$.\\
The first {\em Ansatz} would naturally be to assume
\beq
	H=\sum_{n\in\Z} \left(\alpha_n\hbar^n{\cal H}+\hbar^n\beta_n\right)
	\qquad
	D_i =\sum_{n\in\Z} a_n\hbar^n{\cal D}_i \label{eq:ansatz1}
\eeq
inspired by the lifting of the anomaly presented in \cite{wdef}. A priori, one
could have $a_n$ be a matrix, but the recursion relations arising from this
show that it will have to be proportional to the unit matrix.\\
Inserting (\ref{eq:ansatz1}) into
\beqa
	\left[H(x),H(x')\right]_M &=& i\hbar\left(g^{ij}(x)D_i(x)\delta_{,j}
	(x,x')+(x\leftrightarrow x')\right) \\
	\left[H(x),D_i(x')\right]_M &=& i\hbar H(x)\delta_{,i}(x,x')
\eeqa
we get the following relations
\beqa
	\alpha_k &=& \sum_{n\in\Z} \alpha_n a_{k-n}\\
	a_k &=& \sum_{n\in\Z}\alpha_n\alpha_{k-n}\\
	\beta_k &=& -9\sum_{n\in\Z}\alpha_n\alpha_{k-n-2}
\eeqa
subject to $\alpha_0=a_0=1, \beta_n=0, n\leq 0$. A solution is $\beta_2=-9,
\alpha_0=a_0=1$ all other coefficients vanishing. This is the solution one
would suspect looking at the anomalous bracket. Consequently
\beq
	H(x) = {\cal H}(x)-9\hbar^2
\eeq
and $D_i(x)={\cal D}_i(x)$. The anomaly represents, then, a zero-point
energy or cosmological constant. Inflation can be interpreted as being related
to negative zero-point energy, \cite{casimir}. Thus the form of $H(x)$ suggests
that gravitational fluctuations can inflate and become macroscopic universes
(a big bang scenario). Furthermore, the fact that the negative value is $O(
\hbar^2)$ shows that is a very small quantity and will hence only be important
in the very early universe, and/or at the Planck scale. At larger scales 
(corresponding to later times) it will be negligible.

\subsection{Physical States}
Since the constraints are polynomial
in the phase-space variables the equations defining the physical state
space, $\tilde{\cal C}_{\rm phys}$, become finite order differential
equations. Explicitly, since the constraints are at most quartic in the
phase\-space variables we get
\beqa
	0=[{\cal H},W]_M^+ &=& 2{\cal H}W-\frac{1}{2}\hbar^2\left(E_b^kE_c^l
	\epsilon_a^{~bc}\epsilon^a_{~ef}\frac{\delta^2W}{\delta E^k_e\delta
	E^l_f}+\epsilon_a^{~bc}F_{ij}^a\frac{\delta^2W}{\delta A_i^b
	\delta A_j^c}\right.-\nonumber\\
	&&\hspace{-2cm} \left.2\epsilon_a^{~bc}\left(-\delta^a_e
	(\delta^k_i\partial_j-\delta^k_j
	\partial_i)+\epsilon^a_{~pq}(\delta^p_e\delta^k_iA^q_j
	+\delta^q_e\delta^k_jA_i^p)\right)(\delta^i_l\delta_b^fE^j_c
	+\delta^j_l\delta_c^fE^i_b)\frac{\delta^2W}{\delta E_e^k\delta A^f_l}
	\right)\nonumber\\
	&&+\frac{5}{4}\hbar^4\epsilon_{bc}^{~~a}
	\epsilon_a^{~ef}\frac{\delta^4W}{\delta E^k_e\delta E^l_f\delta A^e_k
	\delta A^f_l} \label{eq:physH}\\
	0=[{\cal D}_i,W]_M^+ &=&2{\cal D}_iW-\frac{1}{2}\hbar^2\left(
	\epsilon^a_{~ef}E^j_a\frac{\delta^2W}{\delta E_e^i\delta E^j_f}-
	\right.\nonumber\\
	&&\left.2\left(-\delta^a_e(\delta^k_i\partial_j-\delta^k_j\delta_i)
	+\epsilon^a_{~mn}(\delta^m_e\delta^k_iA^m_j+\delta^n_e\delta^k_j
	A^m_i)\right)\frac{\delta^2W}{\delta E^k_e\delta A^a_j}\right)
	\label{eq:physD}\\
	0=[{\cal G}_a,W]_M^+ &=& 2{\cal G}_aW + \frac{1}{4}i\hbar^2\delta^j_k
	\epsilon^c_{~ab}\frac{\delta^2W}{\delta A^c_j\delta E^k_b}
	\label{eq:physG}
\eeqa
These coupled equations constitute the equations for the Wigner function
for Ashtekar gravity in vacuum. If we replace ${\cal H}(x)$ by the quantum
Hamiltonian $H(x)$, the only change in these equations will be the appearance
of a $-18\hbar^2 W$ term on the right hand side of (\ref{eq:physH}).
	
\subsection{Loop Formalism and Solutions}
One of the most promising aspects of Ashtekar gravity is the loop transform
of Rovelli and Smolin, \cite{loop}. It is in this formalism that the classical
constraints can be solved seemingly opening up for a consistent quantisation.
Furthermore, the loop states are closely related to Chern-Simons theory
(to be expected considering Witten's expression for knot-invariants) and to
spin networks. Particularly important in the latter case is the appearance of
a quantised spacetime structure in the sense of length, area and volume  
operators with discrete spectra. Consequently, a loop formulation of the
above deformation quantisation is desirable.\\
We will first attempt a direct solution of the equations for a physical state
and hence find $W$, we will then consider the only known solution, namely
the Chern-Simons state $\psi[A]$ and then forms its Wigner function to see
what conditions have to be imposed for this to be a solution.\\
By using the Gauss constraint equation we can rewrite the physicality 
equations as a pair of equations:
\beqa
	0 &=& (2{\cal H}+4i{\cal G}^2)W-\frac{1}{2}\hbar^2\left\{
	i\epsilon_a^{~bc}\epsilon^a_{~gh}E^i_bE^j_c\frac{\delta^2W}{\delta
	E^i_g\delta E^j_h}+\epsilon_a^{~bc}F_{ij}^a\frac{\delta^2W}{\delta
	A_i^b\delta A_j^c}\right\}+\nonumber\\
	&&\qquad\frac{5}{4}\hbar^4\frac{\delta^4W}{\delta A^a_i\delta A^b_j
	\delta E^i_a\delta E^j_b}\\
	0&=& 2F_{ij}^aE^i_aW-\frac{1}{2}\hbar^2\left(i\epsilon^a_{~bc}E^j_a
	\frac{\delta^2W}{\delta E^j_b\delta E^i_c}-2D_i\frac{\delta^2W}
	{\delta E^j_a\delta A^a_j}\right)
\eeqa
We will make the {\em Ansatz}
\beq
	\frac{\delta^2W}{\delta E^i_a\delta E^j_b} = \alpha^{ab}_{ij}W
\eeq
with $\alpha^{ab}_{ij}$ independent of $E$. Inserting this in the two
constraint equations we arrive at
\beq
	\alpha^{ab}_{ij} = 4\hbar^{-2}\left(\delta^{ab}D_iD_j+i\epsilon_c^{~ab}
	F^c_{ij}\right)
\eeq
Thus\footnote{For a Yang-Mills theory only the Gaussian constraints appear and
the corresponding physicality condition can in fact be solved by a similiar
Gaussian Wigner function, namely $W[A,E] \sim \exp(-\alpha c_a^{~bc} F_{ij}^a
E^i_bE^j_c)$ as shown in \cite{def}.}
\beq
	W[A,E] = \exp\left(4\hbar^{-2}\int \delta^{ab}(D_iE^i_a(x))(D_jE^j_b
	(y))+i\epsilon_c^{~ab}\delta(x,y)F^c_{ij}(x)E^i_a(x)E^j_b(y) dxdy+
	...\right)
\eeq
where the remaining terms can be at most linear in $E$, hence we can write
$W$ in symbolic notation as (invoking a generalised summation convention
implying an integration over the continous variables too)
\beq
	W[A,E] = e^{\alpha^{ab}_{ij}[A]E^i_aE^j_b+\beta^a_i[A]E^i_a+\gamma[A]}
\eeq
Now, as was also found for $\alpha$, the coefficients $\beta,\gamma$ can only
depend on $A$ and, moreover, must do so through covariant combinations. This
implies (upto constant terms, which we'll ignore)
\beqa
	\beta^a_i &=& \beta \epsilon_i^{~jk}F_{jk}^a\\
	\gamma &=& \gamma_1 \Tr F^2 + \gamma_2 \Tr \epsilon^{ijk}A_iF_{jk}
\eeqa
where $\beta,\gamma_1,\gamma_2$ are constants. Notice the appearance of the
(three dimensional) Yang-Mills and Chern-Simons actions in $\gamma$. 
Inserting this into the
equations one can collect powers of $E$ to find expressions for the 
coefficients, but as these are rather messy we will not do so here.\\
Notice, furthermore, that this solution gives us directly a Wigner function,
whether this is the Weyl transform of a pure state or not, and if so, what this
pure state is has not been answered. Secondly, we will start from a pure
state, $\psi[A]$, and then construct out of it a Wigner function. We will then
investigate when this state gives rise to a physical solution.\\ 
Consider, then, a connection $A$ and a function $\psi[A]$. The formal Wigner
function is \cite{wigner}
\beq
	W_\psi[A,E] := \int e^{i\int {\rm Tr} B\cdot E d^3x} \bar{\psi}
	[A+\frac{1}{2}B]\psi[A-\frac{1}{2}B]{\cal D}B
\eeq
We will also consider its Fourier transform $E\rightarrow B$, which is just
\beq
	\tilde{W}_\psi[A,B] := \bar{\psi}[A+\frac{1}{2}B]\psi[A-\frac{1}{2}B]
\eeq
The quantities $A,B$ are both Lie algebra valued one forms, and we can perform
a loop transform $A\rightarrow\alpha, B\rightarrow\beta$, where $\alpha,\beta$
are loops. Now, a general function $\psi[A]$ of a connection can be mapped into
a functional of a loop by means of the loop transform 
\cite{loop,loop2,loop3,loop4}
\beq
	\psi[\alpha] := \int \psi[A]h[\alpha,A]{\cal D}A
\eeq
where $h[\alpha,A]$ is the trace of the holonomy (the Wilson loop), i.e.,
\beq
	h[\alpha,A] := {\rm Tr} Pe^{\oint_\alpha A}
\eeq
Doing this we can then define the following functions
\beqa
	W_\psi[\alpha,E] &:=& \int W_\psi[A,E]h[\alpha,A]{\cal D}A\\
	\tilde{W}_\psi[\alpha,B] &:=& \int \tilde{W}_\psi[A,B]h[\alpha,A]
		{\cal D}A \\
	\tilde{W}_\psi[\alpha,\beta] &:=& \int \tilde{W}_\psi[A,B] h[\alpha,A]
	h[\beta,B]{\cal D}A{\cal D}B
\eeqa
We will refer to any of these as a loop transform of the Wigner function 
$W_\psi$. Furthermore, we will usually omit the subscript $\psi$.\\
In terms of loop variables the classical Gauss constraint vanishes identically,
${\cal G}_a=0$. Hence, the quantum modified Gauss constraint equation,
$[{\cal G}_a,W]_M^+=0$, reduces to
\beq
	\epsilon^a_{~bc}B^b_i\hat{\cal T}^i_a\tilde{W}[\alpha,B] =0
\eeq
where we have used
\beqa
	\int\frac{\delta\tilde{W}[A,B]}{\delta A^a_i}h[\alpha,A]{\cal D}A
	&=& - \int\tilde{W}[A,B]\frac{\delta}{\delta A^a_i}h[\alpha,A]
	{\cal D}A\nonumber\\	
	&=& -\int \tilde{W}[A,B]\int ds \delta(x,\alpha(s))\dot{\alpha}^i(s)
	\hat{T}_a(s)h[\alpha,A]{\cal D}A\nonumber\\
	&:=& -\hat{\cal T}_a^i\tilde{W}[\alpha,B]
\eeqa
which follows from the standard formula
\beqa
	\frac{\delta}{\delta A^a_i}h[\alpha,A] &=& \int ds \delta(x,\alpha(s))
	\dot{\alpha}^i(s){\rm Tr}\left( e^{\oint_\alpha A}\tau_a\right)\\
	&:=& \int ds \delta(x,\alpha(s))\dot{\alpha}^i(s)\hat{T}_a(s)
	h[\alpha,A]
\eeqa
it being understood that the Lie algebra generator $\tau_a$ is inserted at the
point $x=\alpha(s)$ along the loop. A possible solution to the quantum Gauss
constraint equation is consequently quite simply
\beq
	\dot{\alpha}^i\perp B^a_i \forall a
\eeq
showing that $B^a_i$ provides a {\em framing} of the loop $\alpha$. It is very
interesting that framing appears naturally in this formalism without the
need of putting it in by hand. In the original relationship between 
Chern-Simons theory and knot invariants, \cite{Witten}, the framing was needed
in order to make the functional integrals convergent. This is a further 
suggestion that deformation quantisation automatically takes care of
regularisation. It is already known that the twisted product can be seen as
regularising the usual one, making products of $\delta$-functions possible.
The possibility of using a Lie algebra-valued one form for framing the loop
was already mentioned in \cite{BFknot}.\\
The next constraint equation to consider is the (spatial) diffeomorphism one,
$o=[{\cal D}_i,W]_M^+$. Performing the Fourier transform $E\rightarrow B$
we get straightforwardly ($\tilde{W}=\tilde{W}[A,B]$)
\beqa
	0 &=& -2iF^a_{ij}\frac{\delta}{\delta B^a_j}\tilde{W}+
	\frac{1}{2}i\hbar^2\epsilon^a_{~bc}B^b_iB^c_j\frac{\delta\tilde{W}}
	{\delta B^a_j}+\nonumber\\
	&&\frac{1}{2}i\hbar^2\left[\delta^a_e(\delta^k_i\partial_j
	-\delta^k_j\partial_i)-\epsilon^a_{~bc}(\delta^b_e\delta^k_i A^c_j
	+\delta^c_e\delta^k_jA^b_i)\right]\left(B^e_k\frac{\delta\tilde{W}}
	{\delta A_j^a}\right)
\eeqa
In order to be able to carry out the loop transform $A\rightarrow\alpha$,
we must then compute the loop transform of $A\tilde{W}, F\tilde{W}$ and
$A\frac{\delta}{\delta A}\tilde{W}$. These are found by brute force
computations.\\
First
\beq
	\int A^a_i\tilde{W}[A,B]h[\alpha,A]{\cal D}A = \frac{1}{2}
	\frac{\delta}{\delta\dot{\alpha}^i}\hat{T}_a\tilde{W}[\alpha,B]
\eeq
From the definition  of $F_{ij}^a$ as the covariant derivative of $A$ we get
\beq
	F^a_{ij}h[\alpha,A] = \left(\partial_i\frac{\delta}{\delta
	\dot{\alpha}^j}\hat{T}^a-\partial_j\frac{\delta}{\delta
	\dot{\alpha}^i}\hat{T}^a+\epsilon^a_{~bc}\hat{T}^b\hat{T}^c
	\frac{\delta^2}{\delta\dot{\alpha}^i\delta\dot{\alpha}^j}
	\right)h
\eeq
This can also be written in terms of the {\em area derivative}, $\Delta_{ij}$, 
as \cite{loop2,loop3}
\beq
	F^a_{ij}h = \Delta_{ij}(s)\hat{T}^a h
\eeq
Putting all of this together we get the following expression for the quantum
diffeomorphism constraint condition ($\tilde{W}=\tilde{W}[\alpha,B]$)
\beqa
	0 &=& 2i\frac{\delta}{\delta B^a_j}\Delta_{ij}\hat{T}^a\tilde{W}
	+\frac{1}{2}i\hbar^2\epsilon^a_{~bc}B^b_iB^c_j\frac{\delta}{\delta
	B^a_j}\tilde{W}-\nonumber\\
	&&\frac{1}{2}i\hbar^2\left(\delta^k_i\partial_j-\delta^k_j\partial_i
	\right)\left(B^a_k\hat{\cal T}^j_a\tilde{W}\right)+\nonumber\\
	&&\frac{1}{4}i\hbar^2\epsilon^a_{~bc}B^e_k\left(\delta^b_e\delta^k_i
	\frac{\delta}{\delta\dot{\alpha}^j}\hat{T}^c+\delta^c_e\delta^k_j
	\frac{\delta}{\delta\dot{\alpha}^i}\hat{T}^b\right)
	\hat{\cal T}^j_a\tilde{W}
\eeqa
Now, the first part of this is the usual Dirac version of the classical
diffeomorphism constraint which is usually solved by demanding that $\tilde{W}
[\alpha,B]$ be a knot invariant, such as the Chern-Simons state, $\psi[A] =
\exp(-iS_{rm CS}[A])$,
\cite{Witten,Nash}. We see that
we have slightly more freedom here. In fact we have two possibilities, either
$\tilde{W}$ is a knot invariant and this equation then gives a condition for
the $B$-dependence of $\tilde{W}$, or $\tilde{W}$ is allowed to be 
non-invariant, or rather a deformed knot-invariant, perhaps defined through
the $q$-deformed Chern-Simons state of \cite{qCS}.\\
The Chern-Simons state leads to
\beqa
	W[A,E] &=& \delta \left(E^a_i-\frac{k}{4\pi}\epsilon_i^{~jk}F^a_{jk}
	\right)\\
	\tilde{W}[A,B] &=& e^{-\frac{ik}{4\pi}\int{\rm Tr}\epsilon^{ijk}B_i
	F_{jk} d^3x} = e^{-iS_{BF}[A,B]}\\
	\tilde{W}[\alpha,B] &=& \langle h[\alpha]\rangle_{\rm BF}
\eeqa
where $S_{BF}$ is the action of the topological BF-theory, and $\langle\cdot
\rangle_{\rm BF}$ denotes the corresponding expectation values. One should
note that $\tilde{W}[\alpha,B]$ is a knot-invariant, since it is diffeomorphism
invariant (coming from a topological field theory) but it is not the usual
Jones polynomial found by Witten, \cite{Witten} (see also \cite{Nash})
\beqs
	\langle h[\alpha]\rangle_{CS} \propto c(k)^{-w(\alpha)}J_q(\alpha)
\eeqs
with $q=\exp\frac{2\pi i}{k+2}$ and $w(\alpha)$ the {\em writhe} of the loop.
That a relationship between BF-theory and knots exists
was shown by Bimonte et al. in \cite{BFknot}.
But the observables which Bimonte et al. show gives the
reciprocal of the Alexander-Conway polynomials is not the Wilson loops which
we consider here.\\
To get a feeling for the nature of these Wigner functions we can
note that it is defined for any Lie algebra, and then consider the simplest
possible example. For a $U(1)$
theory we can actually arrive at an explicit result for $\tilde{W}[\alpha,B]$
and thus see what it looks like, since
\beqas
	\left.\tilde{W}[\alpha,B]\right|_{U(1)} &=& \int e^{-\frac{ik}{4\pi}
	\int \epsilon^{ijk}B_iF_{jk}d^3x+i\int_0^1 \dot{\alpha}^i(s)A_i
	(\alpha(s))ds}{\cal D}A\\
	&=& \int e^{-\frac{ik}{4\pi}\int\epsilon^{ijk}B_iF_{jk}d^3x+i\int
	\Delta^i(\alpha,x)A_i(x)d^3x}{\cal D}A\\
	&=&\delta(\Delta^i(\alpha,x)+\frac{k}{4\pi}\epsilon^{ijk}\partial_jB_k)
\eeqas
Implying the ``Maxwell equation'' $\nabla \times B=-\frac{4\pi}{k}\Delta$, 
i.e., the form factor $\Delta^i$ acts like a source for the ``magnetic'' 
field. The ``physical'' interpretation of the Wigner functions in the
Abelian case is the ``Maxwell'' equations $E=\frac{k}{4\pi}\nabla\times A$ and
$\nabla\times B=-\frac{4\pi}{k}\Delta$. The first imply $\nabla\cdot E:=\rho=0$
whereas the latter imply conservation of the current $j=\Delta, \nabla\cdot 
\Delta=0$. One would expect charges to be located at the ends of curves, 
``field lines'', and since we only deal with loops $\rho=0$. For $SL_2(\C)$
we cannot find such a simple relationship.\\
For the Chern-Simons state, or, rather, the BF-state we can derive a few
simple but useful identities. By direct computation one sees that
\beqas
	\frac{\delta}{\delta A^a_i}\tilde{W}[A,B] &=& -\frac{ik}{4\pi}
		\epsilon^{ijk}(D_kB_j)^a\tilde{W}[A,B]\\
	\frac{\delta}{\delta B^a_i}\tilde{W}[A,B] &=& -\frac{ik}{4\pi}
		\epsilon^{ijk}F_{jk}^a\tilde{W}[A,B]
\eeqas
Inserting the BF-state into the Gauss constraint we get
\beq
	0= -\frac{k}{2\pi}\epsilon^{ijk}(D_iF_{jk}^a)\tilde{W} - 
	\frac{k}{16\pi}\hbar^2\epsilon^{ijk}\epsilon^a_{~bc}D_j(B_k^bB_i^c)
	\tilde{W}
\eeq
the first part, the classical contribution, vanishes by virtue of the Bianchi  
identity, thus confirming that the state is in fact (classically) gauge 
invariant. The second part then gives an equation the $B$-field has to satisfy,
namely
\beq
	\epsilon^{ijk}\epsilon^a_{~bc}D_i(B_j^bB_k^c)=0
\eeq
One particular solution to this is $\epsilon^a_{~bc}B_i^bB_j^c\propto 
F_{ij}^a$. The Bianchi identity then ensures that $\tilde{W}$ satisfies the
quantum Gauss constraint. In any case, we see that this condition is closely
related to the cohomology of the space, since it says that $B_i^bB_j^c
\epsilon^a_{~bc}$ is a Lie algebra valued two form, which is closed with
respect to the covariant derivative. Consequently, if $D^2=0$ then a
solution is for it to be $D$-exact. Now, since $D^2\omega=F\wedge\omega$ for
any $p$-form $\omega$, we see that only flat connections satisfy $D^2=0$. In
that case, locally we have $D\omega=0\Rightarrow \omega=D\chi$, whether this
holds globally depends on the de Rham cohomology of the manifold and on the
Lie algebra cohomology.  Even if $D^2\neq 0$, we can still find a
solution by putting $\epsilon^a_{~bc}B^b_iB^c_j\propto F_{ij}^a$, since the
Bianchi identity then ensures this to be a solution. The solution 
we will find
which also satisfies the spatial diffeomorphism constraint equation will
be precisely of this form.\\
For the spatial diffeomorphism constraint equation we similarly get (after the
Fourier transform $E\rightarrow B$, but before the loop transform $A\rightarrow
\alpha$) by inserting the BF-state
\beqa
	0 &=& -\frac{k}{2\pi}\epsilon^{jkl}\delta_{ab}F_{ij}^aF_{kl}^b+
	\frac{\hbar^2k}{8\pi}\epsilon_{abc}\epsilon^{jkl}B^b_iB^c_jF^a_{kl}
	+\nonumber\\
	&&\frac{\hbar^2k}{8\pi}\left[\delta^a_e(\delta^k_i\partial_j-
	\delta^k_j\partial_i)-\epsilon^a_{~bc}(\delta^b_e\delta^k_iA^c_j
	+\delta^c_e\delta^k_jA^b_i)\right]\left(B^e_k\epsilon^{ljm}(D_mB_l)^a
	\right)
\eeqa
which can be rewritten using ${\rm Tr}\tau_a\tau_b=2\delta_{ab}$ in the
following way
\beqa
	0 &=& -\epsilon^{jkl}{\rm Tr}F_{ij}F_{kl} +\frac{1}{2}\hbar^2
	\epsilon_{abc}\epsilon^{jkl}B^b_iB^c_jF^a_{kl}+\nonumber\\
	&&\frac{1}{4}\hbar^2\epsilon^{ljm}{\rm Tr}\left[(D_iB_j-D_jB_i)(D_mB_l)
	+B_i(D_jD_mB_l)-B_j(D_iD_mB_l)\right]\nonumber\\
\eeqa
Using the Bianchi identity and making the {\em Ansatz} 
\beq
	B_i^aB_j^b\epsilon_{abc} = \alpha F_{ij}^c
\eeq
this reduces to the following conditions
\beqa
	\alpha &=& 4\hbar^{-2}\\
	\Tr\epsilon^{jml}\left(D_j(B_iD_mB_l)-D_i(B_jD_mB_l)\right)&=&0
\eeqa
We will take the latter as a condition on $B$.\\
Moving on, and considering the Hamiltonian constraint we get, after a bit of
algebraic manipulation, the following condition
\beqa
	0 &=& \frac{5k^2}{8\pi^2}(1+\alpha \hbar^2)\epsilon_{abc}\epsilon^{ipq}
	\epsilon^{jrs}F_{ij}^aF_{pq}^bF_{rs}^c+\nonumber\\
	&&\frac{\hbar^2k^2}{32\pi^2}(1+\frac{5}{2}\hbar^2\alpha^{-1})
	\epsilon_{abc}\epsilon^{ipq}\epsilon^{jrs}F_{ij}^a(D_pB_q)^b(D_rB_s)^c
	+\nonumber\\
	&&\frac{3}{2}i\hbar^2k\delta_{ab}\epsilon^{ijk}B_i^aF_{jk}^b+27\hbar^2
\eeqa
To find a solution to this we have to impose extra conditions of $B$ and $F$.
The form of the Hamiltonian constraint equation suggests the 
following condition
\beq
	(D_iB_j)^a=\beta F_{ij}^a
\eeq
(that this is compatible with (117) follows by acting with $\epsilon^{ijk}D_k$
on (117))
and the equation then implies (where we have inserted the expression for
$\alpha$)
\beq
	\beta^2=-20\hbar^{-2}(1+\frac{5}{8}\hbar^4)^{-1}
\eeq 
i.e., $\beta$ must be purely imaginary. The condition then also leads to
(from the last line of (120))
\beq
	\delta_{ab}\epsilon^{ijk} B_i^aF_{jk}^b = i\frac{2}{9}k^{-1}
\eeq
and thus the only states with a finite Wigner functions (i.e., the ones for
which the $BF$-action is finite) are the ones of compact volume.\\
Now, the various conditions on $B$ and $A$ can be summarised in the
following pair of equations
\beqa
	\left[B_i,B_j\right] &=& \frac{\alpha}{\beta} D_iB_j\\
	{\rm Tr}~\epsilon^{ijk}B_iF_{jk} &=& \frac{2i}{9\beta k}
\eeqa
There is an interesting interpretation of this set of equations: The first
states that $B_i$ is a kind of Maurer-Cartan form (on an associated bundle
to the original $su_2$-pincipal bundle) and the second can be rewritten
as ${\rm Tr}\epsilon^{ijk}B_i[B_j,B_k] = const.$ which states that a certain
cocycle (existing, by the way for any Lie algebra) is constant, corresponding
to a notion of some kind of constant volume.\\
Thus the only spacetimes for which the Chern-Simons state is a solution to
this set of quantum deformed physicality conditions is a space where the
``dual'' BF-theory is such that $B$ is an ``associated Maurer-Cartan form'',
which is the ``squareroot of the field strength tensor'' and where the volume
of three-space is finite.\\
We will now turn to the general form of the constraints in the loop formalism.
Performing the transformation $E\rightarrow B, A\rightarrow\alpha$, the
Hamiltonian constraint equation can be written (for a general state, not just
the Chern-Simons/BF-state)
\beqa
	0&=& -2\epsilon_a^{~bc}\frac{\delta^2}{\delta B^b_i\delta B^c_j}
	\Delta_{ij}\hat{T}^a\tilde{W}[\alpha,B]-\nonumber\\
	&&\frac{1}{2}\hbar^2\epsilon_a^{~bc}\Delta_{ij}\hat{T}^a\hat{\cal 
	T}^i_b\hat{\cal T}^j_c\tilde{W}[\alpha,B]-\nonumber\\
	&&\frac{1}{2}\hbar^2\epsilon_{abc}\epsilon^{aef}B^b_kB^c_l
	\frac{\delta^2}{\delta B^e_k\delta B^f_l}\tilde{W}[\alpha,B]+
	\nonumber\\
	&&2\hbar^2\epsilon_a^{~bc}\partial_l\frac{\delta}{\delta B^b_k}\left(
	B^a_k\hat{\cal T}^l_c\tilde{W}\right)+\nonumber\\
	&&2\hbar^2\epsilon_a^{~bc}\epsilon_{bd}^{~~e}\frac{\delta}{\delta 
	B^e_k}\left(B^a_k\frac{\delta}{\delta\dot{\alpha}^l}\hat{T}^d
	\hat{\cal T}^l_c\tilde{W}\right)-\nonumber\\
	&&\frac{5}{4}\hbar^4\epsilon_{bc}^{~~a}\epsilon_a^{~ef}B^e_kB^f_l
	\hat{\cal T}^k_e\hat{\cal T}^l_f\tilde{W}[\alpha,B]
\eeqa
where the last term actually vanishes upon imposing the Gauss condition.
A physical, geometrical interpretation of this equation is difficult to give,
and we will consequently move on.\\
Noting that $B$ is again a one-form, we can perform a second loop transform
$B\rightarrow\beta$, thereby arriving at a Wigner function $\tilde{W}[\alpha,
\beta]$. The Gauss condition then reads
\beq
	\epsilon^a_{~bc}\frac{\delta}{\delta\dot{\beta}^i}\hat{S}^b
	\hat{\cal T}^i_a\tilde{W}[\alpha,\beta]=0
\eeq
where $\hat{S}^b$ is defined in the same way as $\hat{T}^b$ but with the loop
$\beta$ replacing $\alpha$. Similarly the diffeomorphism constraint reads
\beqa
	0 &=& 2\hat{\cal S}^j_b\Delta_{ij}\hat{T}^a\tilde{W}+
	\frac{1}{8}\hbar^2\epsilon^a_{~bc}\frac{\delta^2}{\delta\dot{\beta}^i
	\delta\dot{\beta}^j}\hat{S}^b\hat{S}^c\hat{\cal S}^j_a\tilde{W}
	-\nonumber\\
	&&\frac{1}{4}\hbar^2(\delta^k_i\partial_j-\delta^k_j\partial_i)
	\frac{\delta}{\delta\dot{\beta}^k}\hat{S}^a\hat{\cal T}^j_a\tilde{W}
	+\nonumber\\
	&&\frac{1}{8}\hbar^2\epsilon^a_{~bc}\frac{\delta}{\delta\dot{\beta}^k}
	\hat{S}^e\left(\delta_e^b\delta^k_i\frac{\delta}{\delta\dot{\alpha}^j}
	\hat{T}^c+\delta^c_e\delta^k_j\frac{\delta}{\delta\dot{\alpha}^i}
	\hat{T}^b\right)\tilde{W}
\eeqa
while the Hamiltonian one becomes
\beqa
	0 &=& -2\epsilon_a^{~bc}\hat{\cal S}^i_b\hat{\cal S}^j_c\Delta_{ij}
	\hat{T}^a\tilde{W}-\nonumber\\
	&&\frac{1}{2}\hbar^2\epsilon^{~bc}_a\Delta_{ij}\hat{T}^a\hat{\cal 
	T}^i_b\hat{\cal T}^j_c\tilde{W}-15\hbar^2\tilde{W}-\nonumber\\
	&&\hbar^2\epsilon_a^{~bc}\left[\epsilon_{dc}^{~~a}\hat{\cal S}^k_b
	\frac{\delta}{\delta\dot{\beta}^k}\hat{S}^d\tilde{W}+
	\partial_l\hat{\cal S}^k_b\hat{\cal T}^l_c\frac{\delta}{\delta
	\dot{\beta}^k}\hat{S}^a\tilde{W}+\right.\nonumber\\
	&&\left.\frac{1}{2}\epsilon^a_{~de}\hat{\cal S}^k_b\hat{\cal T}^l_c
	\frac{\delta}{\delta\dot{\alpha}^l}\hat{T}^d\frac{\delta}{\delta
	\dot{\beta}^k}\hat{S}^e\tilde{W}\right]
\eeqa
These conditions are related to the entanglement of the loops (the equations
do not separate, so $\tilde{W}[\alpha,\beta]=\tilde{W}_1[\alpha]\tilde{W}_2
[\beta]$ is not a solution), but I haven't been able to get any further
with them. They are merely included for completeness. An interesting
possibility presents itself though, namely of trying to define a kind of
``regularised'' composition of loops $\alpha \oplus\beta$ such that $\tilde{W}
[\alpha,\beta]=\tilde{W}[\alpha\oplus\beta]$.\\ 
Presumably, one
could gain some insight by using spin networks at this stage. In particular,
there ought to be a close relationship between the formalism proposed here and
the $q$-deformed spin networks introduced by Major and Smolin, \cite{qspin},
since the coupling constant in the Chern-Simons state is left arbitrary in
our approach, and it is this coupling constant which provides the quantum
deformation in the paper by Major and Smolin. But all of that is
beyond the scope of the present paper.

\section{Ashtekar Gravity Lower Dimensions}
In order to shed some light on the meaning of this formalism we will briefly
study the deformation quantisation of gravity in $d=2+1$ and $d=1+1$
dimensions. Now, these theories are classical trivial, the Einstein equations 
amounting to Ricci
flatness which in three dimensions imply true flatness (i.e., vanishing of the
entire curvature tensor), and an empty set fo equations in two dimensions. 
Thus ``gravity'' in $d\leq 3$ is a theory without
local physical degrees of freedom. Most of the insights already gained are
specific to the proper physical dimensionality of four, but the topological
aspects will turn up much clearer in lower dimensions.\\
As already mentioned, the constraints in $d=2+1$ are \cite{Ashtekar2}
\beqa
	\epsilon^{ij}D_iE_j^a &=& 0\\
	F_{ij}^a &=& 0
\eeqa
which precisely state that the space is Ricci flat, and if the metric is 
non-singular, the space is completely flat. Consequently, we will expect the
metric in a quantum theory to be trivial except in a number of isolated points
where the metric is singular. Note, furthermore, that the field $E_i^a$ is {\em
not} a zweibein/dyad but a $SO(2,1)$-valued vector field ($i=1,2$ but 
$a=1,2,3$).\\
The conditions for a state $W$ to be physical turn out to be
\beqa
	0 &=& [\epsilon^{ij}D_iE_j^a, W]_M^+ = 2\epsilon^{ij}D_iE_j^a+
	\frac{1}{2}i\hbar^2\epsilon^{ij}\epsilon^a_{bc}\frac{\delta^2W}
	{\delta A^b_i\delta E^b_j}\\
	0 &=& [F_{ij}^a,W]_M^+ = 2F_{ij}^aW -\frac{1}{2}i\hbar^2g_{kj}g_{li}
	\epsilon^{abc}\frac{\delta^2W}{\delta A^b_k\delta E^c_l}
\eeqa
In a loop transformed formulation the first will again give the $B$-field
Fourier-dual to $E$ to be framing the loop. The only difference from
$d=3+1$ is that the condition will now be $B_i^a\dot{\alpha}_j\epsilon^{ij}=0$
and not $B^a\perp\dot{\alpha}$. In fact the loop plus BF formulation (i.e.,
$A\rightarrow\alpha, E\rightarrow B$) of these constraints is
\beqa
	0 &=& \epsilon^{ij}\epsilon^a_{bc}B_i^b\hat{\cal T}_j^c\tilde{W}\\
	0 &=& 2\Delta_{ij}\hat{T}^a\tilde{W}+\frac{1}{2}\hbar^2\epsilon^a_{bc}
	B_j^b\hat{\cal T}^c_i\tilde{W}
\eeqa
which can be reduced to simply (by contracting with $\epsilon^{ij}$)
\beq
	\Delta_{ij}\hat{T}^a\tilde{W} =0
\eeq
stating that $\tilde{W}$ is a diffeomorphism invariant of a framed loop.\\
Loops in two dimensions, i.e., on some surface of genus $g$, can only depend
on the homotopy class of the loop, suggesting that $\tilde{W}$ is a homotopy
invariant. Consequently, $\tilde{W}$ can only depend on a loop $\alpha$ through
its winding numbers around the $g$ different holes in the surface, i.e.,
$\tilde{W}$ depends on $g$ integers $n_1,...,n_g$ which are the winding numbers
of the loop $\alpha$. This is also what the classical analysis shows,
\cite{Ashtekar2}, but this is hardly surprising since we have just seen that
the quantum constraints reduce to their classical counterparts, plus a
relationship between the field $B$ and the loop. The only
possible extra quantum modification is $E_i^a$, and hence the metric $g_{ij}$,
to be singular at a finite number of points. In that case $\tilde{W}$ will also
depend on the residues at these points. In a two-loop formalism (i.e., $B
\rightarrow\beta$), the Wigner function can only depend on the intersection
number of the two loops besides their homotopy class.\\
We cannot carry the analogy with $d=3+1$ gravity any further since the 
Chern-Simons state does not exist in two dimensions. The BF-state does, but
the $B$ field is then a zero-form an thus not related at all to the electric 
field $E$ on the $2d$ surface. In the full $2+1$ dimensional spacetime
manifold one can take $B\sim E$ to get $S_{BF}\sim S_{EH}$ where $S_{EH}$ is
the Einstein-Hilbert (or rather Palatini) action, which is itself a
BF-theory in $d=2+1$ dimensions.\\
For $d=1+1$, the Lorentz algebra becomes the Abelian algebra $so(1,1)$, hence
the constraints becomes simply 
\beq
	\partial E = 0\qquad F=\partial A=0
\eeq
which implies that the quantum physicality conditions reduces to their
classical part
\beq
	W\propto\delta(\partial E)\delta(\partial A)
\eeq
Thus, the theory is completely trivial classically as well as quantum
theoretically and is not worth spending any more time on.

\section{Conclusion}
We have seen that a deformation quantisation of gravity is possible, although
anomalies turn up in as well the ADM as the Ashtekar formulation. In the
latter, however, the anomaly is merely a central extension and hence
liftable. In any case, the presence of an anomaly signals the breakdown of
diffeomorphism invariance. This can either imply (1) the presence of a
non-vanishing zero point energy, or (2) the appearance of a scale below
which classical gravitation fails. In any case, it shows that the ``time
evolution'' constraint, the Hamiltonian one, is no longer described by a
scalar on the spatial hypersurface $\Sigma$. Hence, the quantum version of
it must contain {\em some} information which the classical doesn't -- this
could be a preferred direction, a scale or an origin. In the first case, we
would expect the constraint to be vector-like, but that does not seem to be
the case.\\
We showed that a solution could be found by assuming a Chern-Simons
state $\psi[A]$ (which then gave rise to a BF-state) even for $\Lambda=0$, but
only if the $B$-field of the corresponding BF-theory was an ``associated
Maurer-Cartan form'' and if the volume of three-space was finite.\\
In the general
loop formalism, the field $B$ -- the Fourier transformed of the dreibein --
became related to the imbedding of the loop. In a two-loop formalism this
were formulated as the necessity of entanglement of the two loops. One should
also note that for the Chern-Simons state, the formal Wigner function becomes
a knot invariant, closely related to the usual Jones polynomial. It is also
worth noticing that framing of loops appeared naturally, was in fact imposed
by the quantum modified Gauss constraint, in this formalism, and didn't have
to be introduced by hand in order to give well-defined expectation values.

\end{document}